\documentclass{PoS}
\DeclareMathAlphabet{\altmathcal}{OMS}{cmsy}{m}{n}
\DeclareSymbolFontAlphabet{\mathnormal}{letters}
\DeclareMathAlphabet{\mathcal}{OMS}{cmsy}{m}{n}
\usepackage{xfrac}
\usepackage{upgreek}

\newsavebox{\twosubbox}
\usepackage{graphicx}
\usepackage{subcaption}
\title{The Spectrum of the Axion Dark Sector, Cosmological Observable and Black Hole Superradiance Constraints}

\ShortTitle{Constraints on the number of axion-like fields}

\author{\speaker{Matthew J. Stott}\thanks{Based on work in collaboration with David J. E. Marsh \cite{Stott:2018opm} (Instit\"{u}t f\"{u}r Astrophysik, G\"{o}ttingen).}\\
        Theoretical Particle Physics and Cosmology Group, Department of Physics,\\ King's College London, University of London, Strand, London, WC2R 2LS, United Kingdom \\
        E-mail: \email{matthew.stott@kcl.ac.uk}}

\abstract{Consistent frameworks of quantum gravity often predict the existence of large numbers of ultralight pseudoscalar degrees of freedom, forming the phenomenological landscape of the String Axiverse. The complexity of the extra-dimensional compactification manifolds and vacua determine that these fields could possess parameters with cosmologically significant scales, which span many decades. Astrophysical observations of stellar binary and supermassive black hole systems can be used to exclude the existence of certain ultralight massive bosons, via the superradiance phenomenon. In this work it is shown how these measurements can be used to constrain properties of statistical distributions for the masses of multiple bosonic field theories, inspired by axion field alignment models and an explicit realisation of the string axiverse in M-theory. Such a methodology can exclude $N_{\rm ax} \geq 30$ axion-like fields with a range of mass distribution widths and central values spanning many orders of magnitude, covering axion phenomenologies important to the dark sector and grand unified theories. This is demonstrated for several examples of axions in string theory and M-theory, where the mass distributions in certain cases take universal forms found in random matrix theory.}

\FullConference{The 39th International Conference on High Energy Physics (ICHEP2018)\\
		4-11 July, 2018\\
		Seoul, Korea}

\begin{document}

\section{Axions and the String Axiverse}

Axions or axion like particles (ALPs) are omnipresent components in string/M-theory frameworks which serve as consistent theories of quantum gravity \cite{Arvanitaki:2009fg, Svrcek:2006yi}. The complex topological properties of the extra dimensional manifolds are often associated with a large number of moduli possessing scales likely important to understanding the full evolution of the cosmic history. A full and detailed understanding of the systematics of the full scalar potential is an enormously imposing task and so various low energy descriptions of these models have enjoyed the benefits of effective field theories (EFTs), incorporating random matrix theory (RMT) and the universal nature their cosmologically defining parameters could take. If these parameters, the axion decay constant, $f_a$ or field mass, $\mu_{\rm ax}$ span specific regions then ALPs can contribute to either the dark matter or dark energy densities of the Universe at the current epoch.  ALPs with a Compton wavelength commensurate to that of spinning astrophysical black holes (BHs) predict a resonant susceptibility to the Penrose process, spinning down BHs via superradiance. Observed masses of stellar and supermassive BHs (SMBHs) give a probeable axion mass window, $10^{-20}\ {\rm eV}\lesssim \mu_{\rm ax} \lesssim 10^{-11}\ {\rm eV}$ providing a powerful methodology to place model independent constraints on both the mass spectrum and the allowed number of fields, $N_{\rm ax}$.  See Sections IV and V of Ref.~\cite{Stott:2017hvl} for constraints on the dark sector cosmology of the RMT axiverse models outlined in Section~\ref{sec:effective_theory}. 
\section{The Effective Random Matrix Axion Mass Spectrum}
\label{sec:effective_theory}
\begin{figure}
\begin{center}
\resizebox{.99\textwidth}{!}{%
\includegraphics[height=3cm]{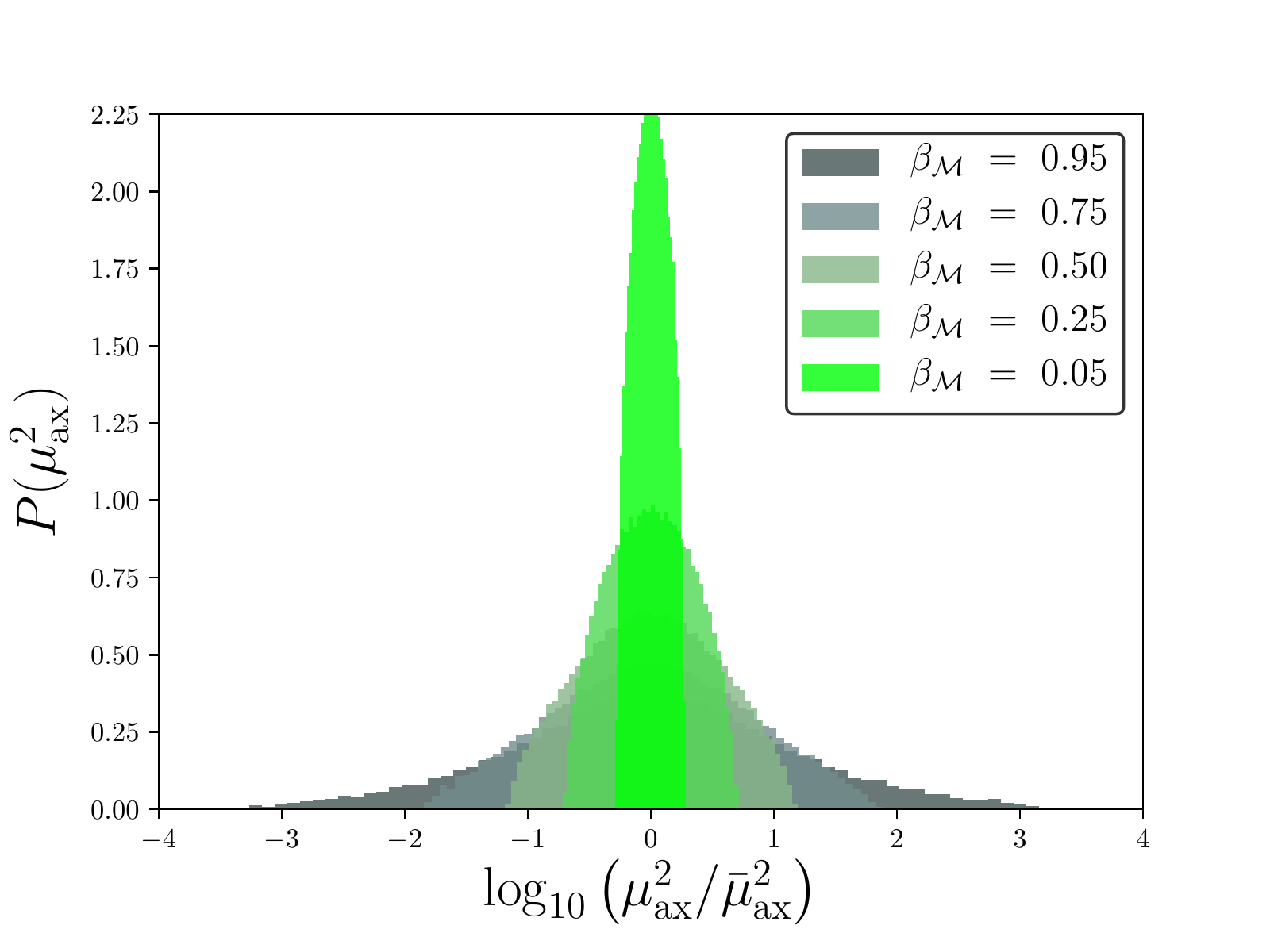}%
\quad
\includegraphics[height=3cm]{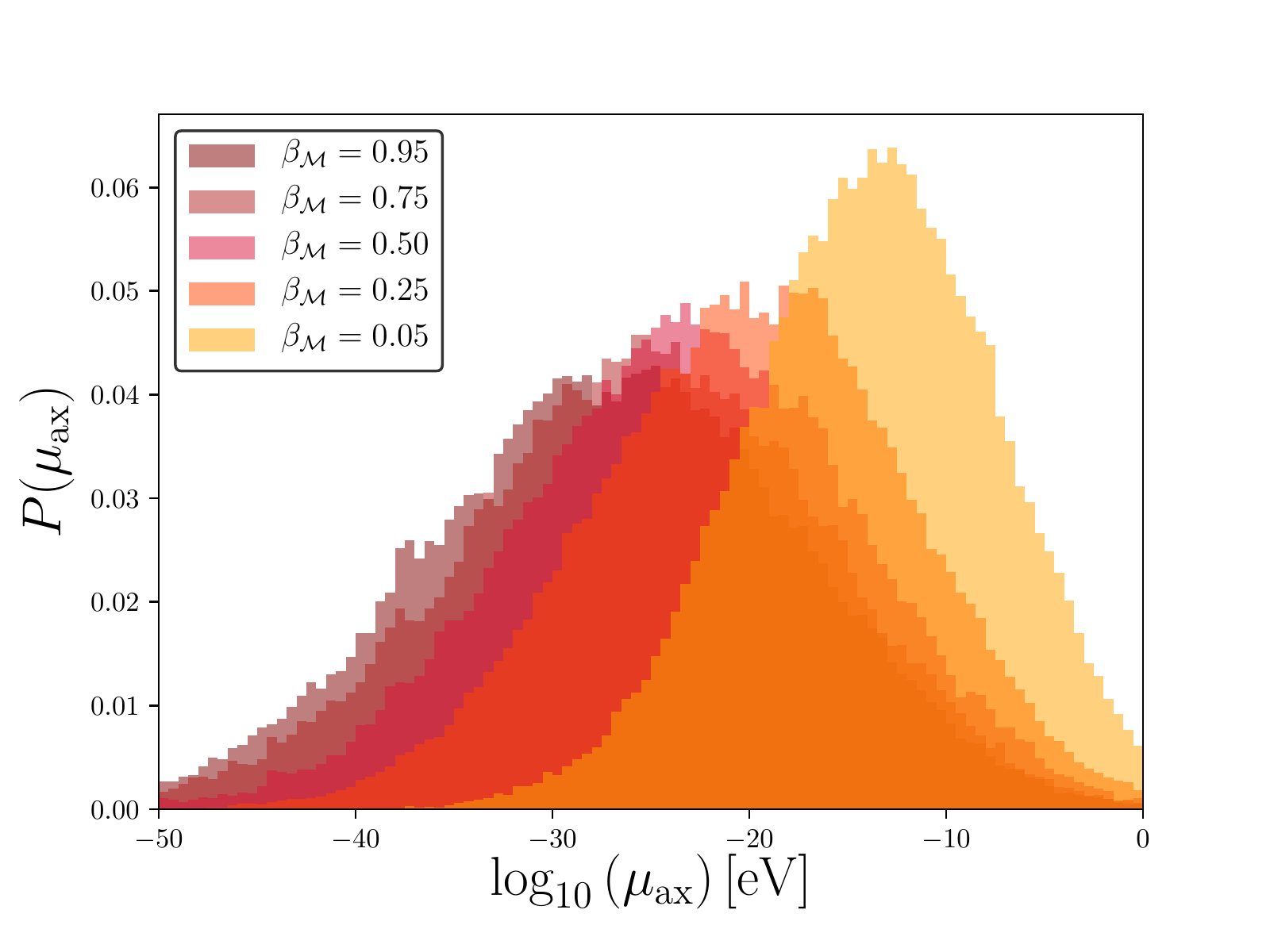}%
\quad
\includegraphics[height=3cm]{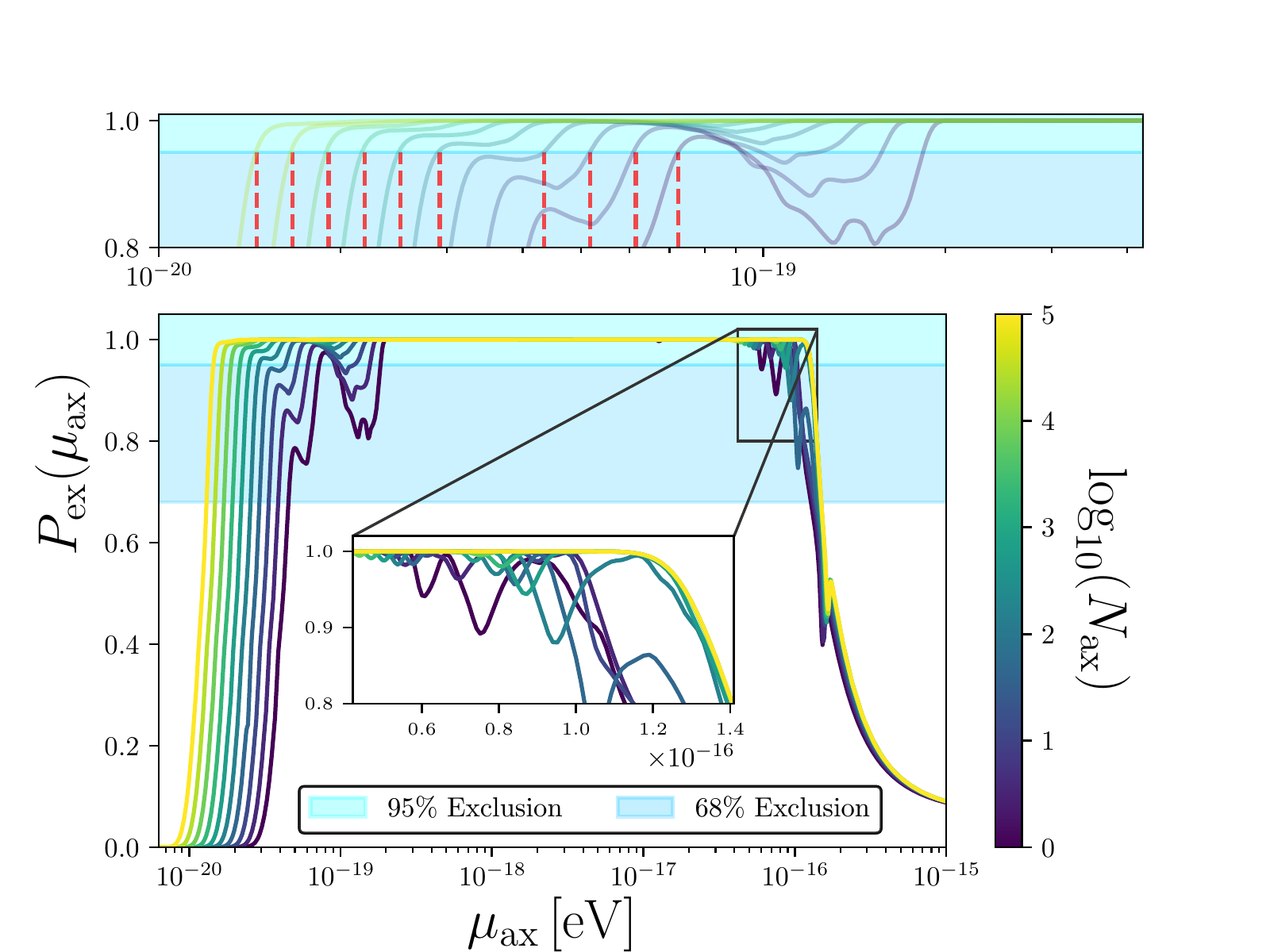}%
}
\end{center}
\caption{\emph{Left panel:} Normalised eigenvalue spectra for axion masses in the Wishart class alignment model in Section~IIIC2 of Ref.~\cite{Stott:2017hvl}. \emph{Middle panel:} M-theory eigenvalue spectra for axions masses using a fixed value for the average three-cycle volume $\langle V_X \rangle = 25 \approx \sfrac{1}{\alpha_{\rm GUT}}$, in Section~IIID of  Ref.~\cite{Stott:2017hvl}. \emph{Right panel:} $P_{\rm ex} \left( \mu_{\rm ax} \right)$ exclusion functions for degenerate mass axion populations. }
\label{fig:mass_spectra}
\end{figure}

The most general effective action for ALPs is such that the kinetic matrix, ${\mathcal{K}_{ij}}$ related to the K\"{a}hler metric in supersymmetric theories is both non-diagonal and non-canonically normalised. The generic multi axion Lagrangian takes the form, 
\begin{equation}
\mathcal{L}_{\rm ax} = -\sum_{i,j = 1}^{N_{\rm ax}}\mathcal{K}_{ij} \partial_{\nu} \theta_i \partial^{\nu} \theta_j - \sum_{\alpha = 1}^{n_{\rm inst}}\sum_{j = 1}^{N_{\rm ax}}\Lambda_\alpha U_\alpha(\mathcal{Q}_{j,\alpha}\theta_j + \delta_\alpha),	
\label{eq:multiaxion}
\end{equation}
where $\theta_i$ are the dimensionless axion fields and $U$ is a general periodic instanton potential with charge matrix, $\mathcal{Q}$ and phases, $\delta$. An expansion to quadratic order of the local minima associated with well-aligned axion potentials in the limit $n_{\rm inst} = N_{\rm ax}$ reveals a spectrum of fields retaining discrete shift symmetries, encoded in a mass matrix, $\mathcal{M}_{ij}$. A synchronous diagonalisation and rescaling of the fields in canonical coordinates reduces Eq.~(\ref{eq:multiaxion}) to the form, 
\begin{equation}
\mathcal{L}_{\rm ax} = -\frac{1}{2}\partial_{\nu}\phi_i\partial^{\nu}\phi_i - \frac{1}{2}{\rm diag}(\mu^2_{\rm ax})\phi_i \phi_i\,.
\label{eq:finall}
\end{equation}
The model's spectrum is given by the mass eigenstates, $\{\mu_i\}$, who's spectral limits can be derived from the universal nature of RMT. It has been shown that a series of RMT models based on axion field alignment and an explicit realisation of the string axiverse in $G_2$ compactified M-theory \cite{Acharya:2010zx} give rise to a spectrum of fields with mass eigenstates well approximated by a log-normal distribution \cite{Stott:2017hvl}. Example spectra are displayed in the \emph{left} and \emph{middle panels} of Fig.~\ref{fig:mass_spectra}, with universal forms determined by the powerful principles in multivariate statistics akin to the multiplicative central limit theorem. Each sample covariance matrix is constructed in the Wishart form, $\mathcal{K}_{ij},\ \mathcal{M}_{ij} = N^{-1} Y_{ik}^T Y_{kj} $ where the dimensions of the sub-matrices are controlled by a shaping parameter, $\beta_{\mathcal{M}}\in (0,1]$ (see Section III of Ref.~\cite{Stott:2017hvl}). The following summarises derived constraints from the more detailed work in Ref.~\cite{Stott:2018opm}.

\section{Black Hole Superradiance Constraints on the Number of Axions}

\begin{figure}
\begin{center}
\resizebox{.99\textwidth}{!}{%
\includegraphics[height=3cm]{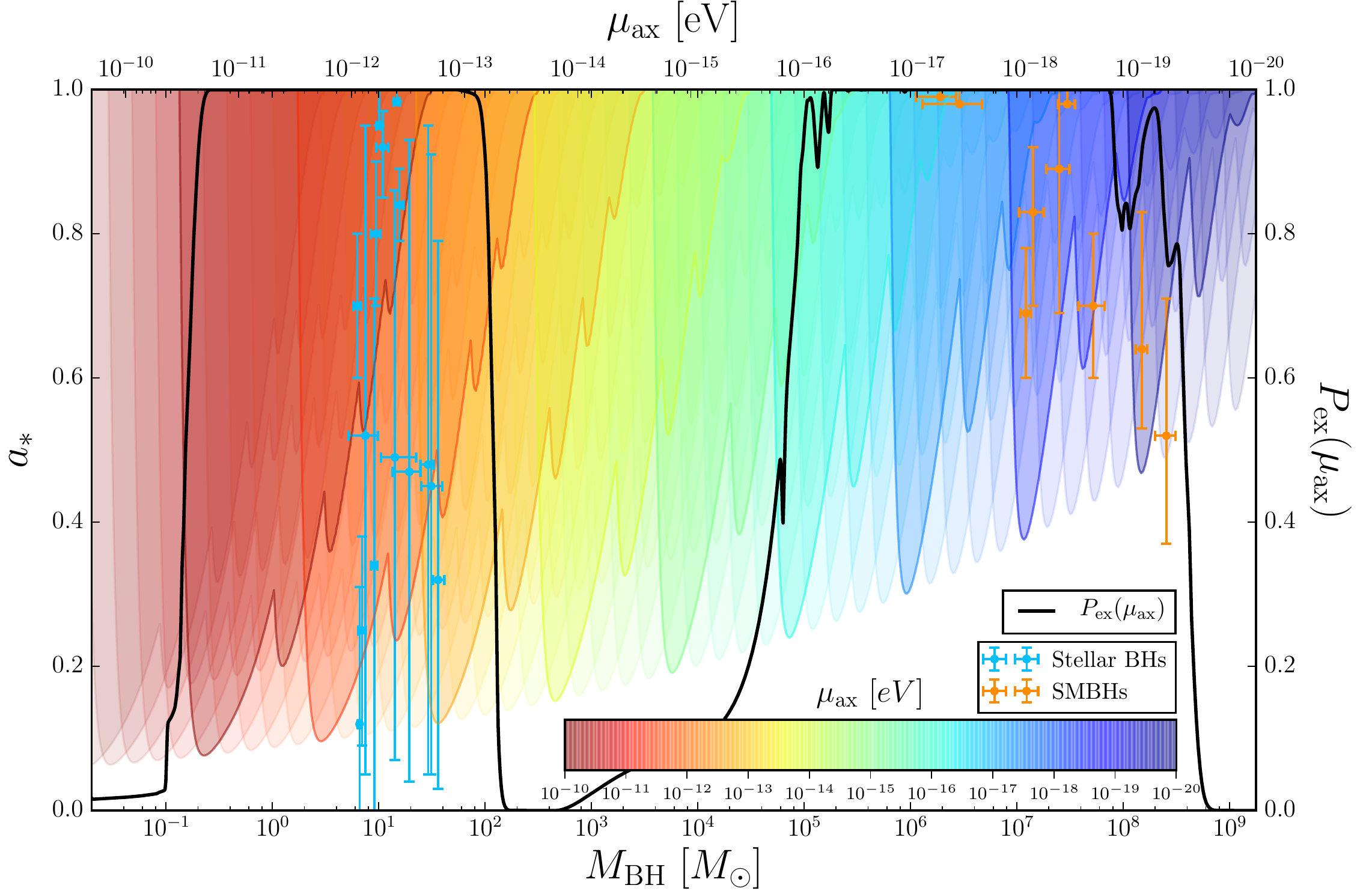}%
\quad
\includegraphics[height=3cm]{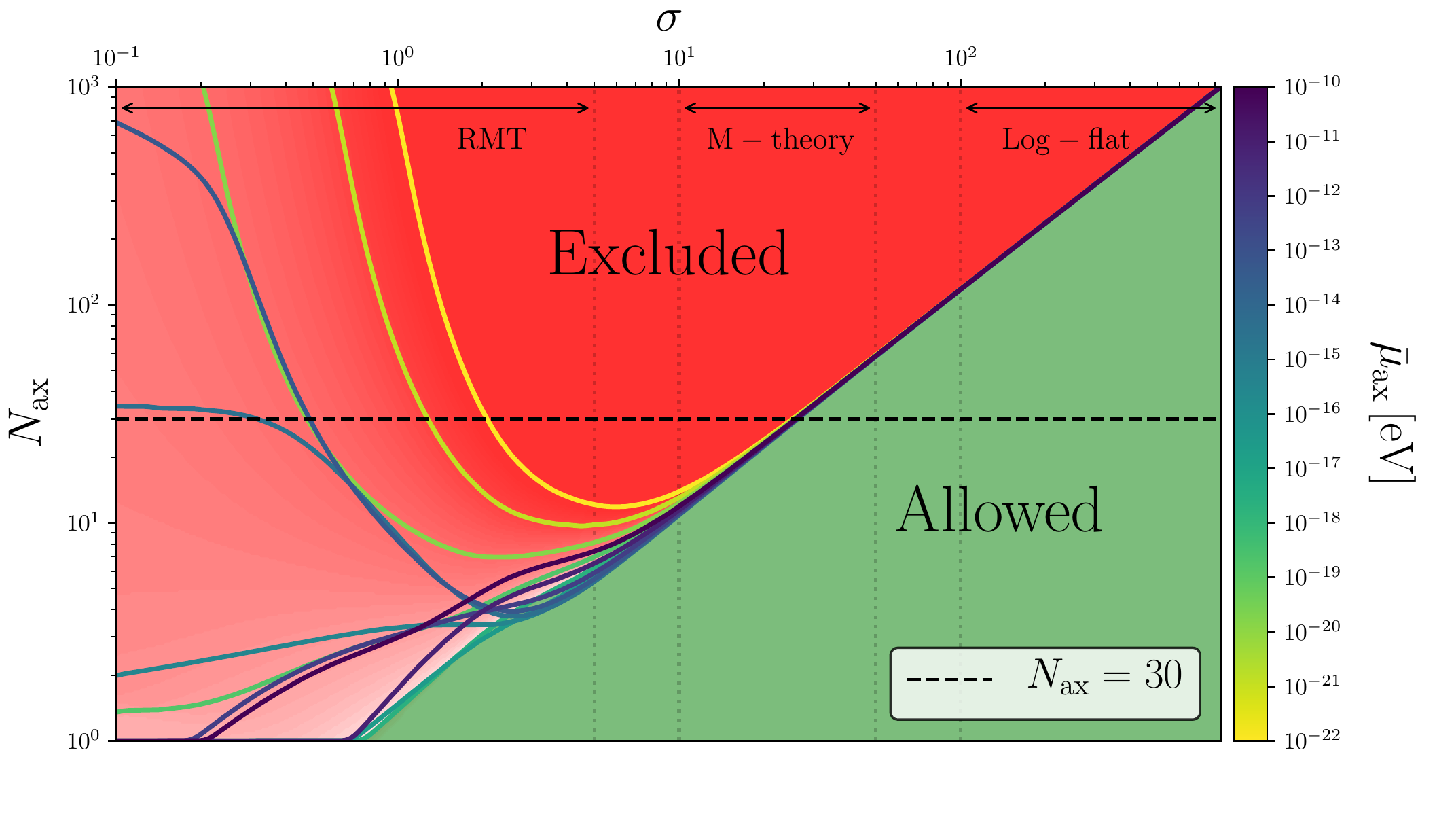}%
}
\end{center}
\caption{\emph{Left panel:} Isocontour exclusion bound thresholds for time scales, $\tau_{\rm Salpeter} \simeq 4.5 \times 10^7 \ {\rm yrs} $ for each of the dominant orbital/azimuthal quantum modes, along with the calculated total exclusion probability function (\emph{black} line) using xray/BH binary and SMBH systems (\emph{blue} and \emph{orange} data points). \emph{Right panel:}  95\% exclusion isocontours regions for log-normal axion mass distributions as a function of the width, $\sigma$, and number of fields, $N_{\rm ax}$, for various central masses, $\bar{\mu}_{\rm ax}$.}
\label{fig:bh_data}
\end{figure}

ALPs in the presence of a rotating BH, conforming to the Kerr geometry, form a scalar-BH condensate system \footnote{The theoretic foundations of BH superradiance have been extensively covered in Refs. \cite{Arvanitaki:2010sy,Brito:2015oca}}. The following constraints hold in the limit of small field self-interactions $\left( f_a \gtrsim 10^{14}\  {\rm GeV} \right)$ where the evolution of each field is independent on the Kerr background and their effects are approximated by a linear summation over the axion population size, $N_{\rm ax}$.  The general condition for mode amplification of the scalar condensate follows the superradiance condition, $\sfrac{\omega}{m}<\omega_+$. An intrinsic prediction of superradiant instabilities associated with bosonic fields, which follow this condition, is the presence of isocontour exclusion regions in the BH Mass-Spin Regge plane. Estimates of the relevant instability time scales, $\uptau_{\rm SR}$  partnered with accurate and reliable spin measurements for BHs, can be used to impose meticulous constraints on the allowed masses the fields can take, inside the region $7\times 10^{-20}\text{ eV}\leq \mu_{\rm ax}\leq 1\times 10^{-16}\text{ eV}$ and $7\times 10^{-14}\text{ eV}\leq \mu_{\rm ax} \leq 2\times 10^{-11}\text{ eV}\,$ as shown in the \emph{left panel} of Fig.~\ref{fig:bh_data}.
It is also possible to constrain a population of non-degenerate masses by first considering the degenerate case, as it is trivial to treat for any number of fields whereby the superradiance rate, additive in $N_{\rm ax}$, defines the total rate, $\Gamma_{\rm tot}=N_{\rm ax}\Gamma_{\rm SR}$. The \emph{right panel} of Fig.~\ref{fig:mass_spectra} shows the 95\% excluded regions for a degenerate population of masses, $\mu_{\rm ax}$. The relevant exclusion bounds change by less than an order of magnitude compared to the single field case for $N_{\rm ax}\lesssim 10^5$, ensuring an increase in the superradiance rate for multiple fields can be neglected when computing the total exclusion probability function. With the effect of rate addition neglected, the exclusion probability for a mass distribution is trivial to construct from the single field overlap integral. The general probability that a given model's parameters, $\theta$ are allowed for $N_{\rm ax}$ becomes, $P_{\rm al}(\theta,N_{\rm ax}|\mathcal{M}) = \left[\int {\rm d}\mu_{\rm ax}\,\, p(\mu_{\rm ax} |\theta,\mathcal{M}) P_{\rm al}(\mu_{\rm ax}|N_{\rm ax}=1) \right]^{N_{\rm ax}}$. The allowed number of ALPs drawn from log-normal distributions as a function of the width, $\sigma$ and central value, $\bar{\mu}_{\rm ax}$ are detailed in the \emph{right panel} of  Fig.~\ref{fig:bh_data}, where regions above the contours are excluded. Certain ranges of $\sigma$ correspond closely to RMT and M-theory mass spectra. For $1\lesssim\sigma\lesssim 20$, $N_{\rm ax}\geq 30$ is excluded for an extremely wide range of central masses important to axion cosmology.

It has been shown BH superradiance can place strong constraints on the possible existence of a spectrum of ultralight ALPs with small self-interactions by considering the central masses and population numbers for universal forms of the fields spectra. It is expected that their masses should follow such particular statistical distributions independently of the microscopic details of the compact space. In particular the benchmark value of $N_{\rm ax} \approx 30$ found in the majority of known Calabi-Yau manifolds can be excluded for a wide range of distribution parameters.

 \section*{Acknowledgments}
M.J.S is grateful to the ICHEP 2018 organising committee for their invitation for participation and for awarding the registration fee grant along with the Centre for Doctoral Studies and King's College London for supporting with travel costs by awarding the Conference Fund Grant. This research is supported by funding from the UK Science and Technology Facilities Council (STFC).

\end{document}